\documentclass[reprint,aps,english,prx,floatfix,superscriptaddress]{revtex4-1}

\usepackage[colorlinks=true,urlcolor=blue,citecolor=blue,linkcolor=blue]{hyperref}
\usepackage[all]{hypcap}

\usepackage{amssymb}
\usepackage{graphicx}
\usepackage{amsmath,color}
\usepackage{color}
\usepackage{blkarray}
\usepackage{braket}

\DeclareMathOperator{\Tr}{Tr}

\newcommand{\eqnref}[1]{(\ref{#1})}

\newtheorem{algorithm}{Algorithm}

\newcommand{\rlll}{\rangle\langle}

\makeatletter

\begin{document}

\title{Extracting entanglement geometry from quantum states}

\author{Katharine Hyatt}
\affiliation{Department of Physics, University of California, Santa Barbara, California 93106, USA}

\author{James R. Garrison}
\affiliation{Joint Quantum Institute and Joint Center for Quantum Information and Computer Science, National Institute of Standards and Technology and University of Maryland, College Park, Maryland 20742, USA}
\affiliation{Department of Physics, University of California, Santa Barbara, California 93106, USA}

\author{Bela Bauer}
\affiliation{Station Q, Microsoft Research, Santa Barbara, California 93106, USA}

\begin{abstract}
Tensor networks impose a notion of geometry on the entanglement of a quantum system.
In some cases, this geometry is found to reproduce key properties of holographic dualities, and subsequently much work
has focused on using tensor networks as tractable models for holographic dualities.
Conventionally, the structure of the network -- and hence the geometry -- is largely fixed \textit{a priori} by the choice of tensor
network ansatz. Here, we evade this restriction and describe an unbiased approach that allows us to extract the appropriate geometry from a given quantum state.
We develop an algorithm that iteratively finds a unitary circuit that transforms a given quantum state into an unentangled product state. We then
analyze the structure of the resulting unitary circuits. In the case of non-interacting, critical systems in one dimension,
we recover signatures of scale invariance in the unitary network, and we show that appropriately defined geodesic paths between
physical degrees of freedom exhibit known properties of a hyperbolic geometry.

\end{abstract}

\maketitle

Tensor networks have proven to be a powerful and universal tool to describe quantum states. Originating
as variational ansatz states for low-dimensional quantum systems, they have become a common language between condensed
matter and quantum information theory.
More recently, the realization that some key properties of holographic dualities~\cite{thooft1993,susskind1995,Maldacena1999,witten1998,Aharony2000}
are reproduced in certain classes of tensor network states (TNS)~\cite{swingle2012,evenbly2011} has
led to new connections to quantum gravity. In particular, many
questions about holographic dualities appear more tractable in TN
models~\cite{qi2013,beny2013,Mollabashi2014,pastawski2015,Bao2015,Miyaji2015,Czech2016,hayden2016,yang2016,kehrein2017,qi2017}.
The study of the geometry of TN states underlies these developments. Here, the physical legs of the
network represent the boundary of some emergent ``holographic'' space that is occupied by the TN. While in
networks such as matrix-product states (MPS)~\cite{Fannes1992,white1992,ostlund1995} and projected entangled-pair states (PEPS)~\cite{gendiar2003,nishino2001,verstraete2004}
this space just reflects the physical geometry, other networks -- such as the multi-scale
entanglement renormalization ansatz (MERA)~\cite{vidal2007,vidal2008} --
can have non-trivial geometry in this space~\cite{evenbly2011}. We will refer to this geometry as ``entanglement geometry''.

\begin{figure}
  \includegraphics{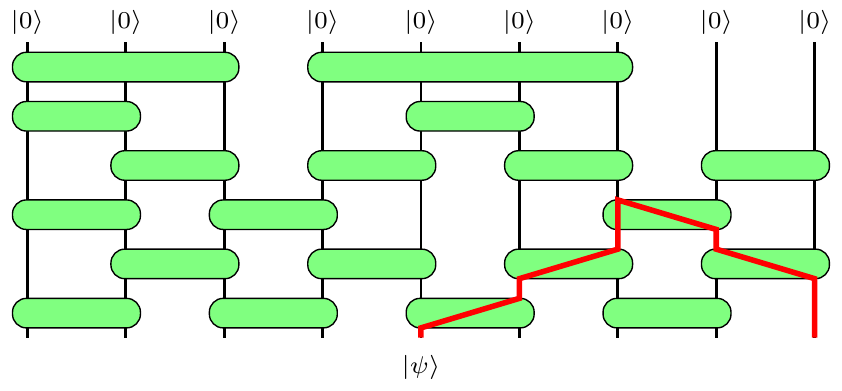}
  \caption{Example of a two-local unitary circuit, where each unitary acts only on the two qubits that are at its ends.
  The thick red line indicates a geodesic between the 5th and 9th qubit (from the left), following a path through the circuit as given by Fig.~\ref{fig:length}.
  \label{fig:circuit} }
\end{figure}

In this paper, we investigate whether this entanglement geometry can be extracted from a given quantum state without pre-imposing
a particular structure on the TN~\cite{Cao2017}. We first describe a greedy, iterative algorithm that, given a
quantum state, finds a 2-local unitary circuit that transforms this state into an unentangled (product) state (see Fig.~\ref{fig:circuit}).
Such circuits, composed from unitary operators acting on two sites (which are not necessarily
spatially close to each other), can be viewed as a particular class of TNS where the tensors are the unitary operators
that form the circuit.

We then develop a framework for analyzing the geometry of these circuits. First, we introduce a locally computable notion of distance
between two points in the circuit, thus inducing a geometry in the bulk. We then focus on a particular property of this geometry, 
the length of geodesics (shortest paths through the circuit) between physical (boundary) sites.
A similar quantity has been previously discussed as a diagnostic of geometry in tensor networks~\cite{evenbly2011},
and reveals similar information as the minimal spanning surface in the celebrated Ryu-Takayanagi (RT) formula for the entanglement entropy in
AdS/CFT~\cite{ryu2006,ryu2006-long}.
Crucially, our definition takes into account the strength of each local tensor, and thus allows us to numerically compute
an appropriate length without imposing additional restrictions on the tensors~\cite{pastawski2015} or \textit{a priori} knowledge of the emergent geometry.

Applying these techniques to many-particle quantum states, we observe three regimes:
(i) a flat (zero curvature) two-dimensional geometry,
(ii) a hyperbolic two-dimensional geometry,
and (iii) a geometry where the geodesic distance between all points is equal,
which corresponds to zero (fractal) dimension.
We first observe these in eigenstates of non-interacting fermions in a disorder potential. For low-energy eigenstates with weak disorder,
we find a hyperbolic geometry and thus recover key aspects of the AdS/CFT duality~\cite{thooft1993,susskind1995,Maldacena1999,witten1998,Aharony2000}.
Going beyond eigenstates, we study a quench from the localized to the delocalized regime,
i.e.\ the evolution of a localized initial state under a Hamiltonian with vanishing disorder potential. In this case, the geodesics reveal detailed information
about the deformation of the emergent geometry, which progresses from flat geometry (i) to zero-dimensional (iii).
This process reproduces certain aspects of previous holographic analyses of quantum quenches~\cite{hubeny2007,Abajo-Arrastia2010,Sonner2015}.

In a complementary approach discussed in App.~\ref{sec:lightcone}, we also examine the nature of emergent light cones in the unitary network. In the case of critical systems, these are found
to exhibit features of scale invariance.
In the cases of localized and thermal states, the light cones reveal that the entanglement is fully encoded in local and global operators, respectively.

\emph{Disentangling algorithm---}
Our algorithm for finding a unitary disentangling circuit is in many ways inspired
by the strong-disorder renormalization group~\cite{ma1979,fisher1994}. However, there are two crucial differences.
First, instead of acting on the Hamiltonian, the algorithm acts on a particular state. Second, rather than on the energetically
strongest bond, at each step the algorithm works on the most strongly entangled pair of sites. The algorithm has two desirable properties.
First, it works for a broad class of input states, including states that have area law and volume law entanglement.
This comes at the cost of generating circuits that cannot in general be contracted in polynomial time.
Second, each iteration of the iterative algorithm is completely determined by the output of the previous iteration; we thus
avoid solving the challenging non-linear optimization problems that are usually encountered when optimizing a tensor
network.
Similar algorithms have been put forward in Refs.~\onlinecite{chamon2014, kehrein2017}.

We take as input a quantum state $|\psi\rangle$ on a lattice $\mathcal{L}$.
We denote as $\rho_{ij}$ the reduced density matrix on sites $i, j \in \mathcal{L}$, $\rho_{ij} = \Tr_{\mathcal{L}\setminus \{i,j\} } |\psi\rangle\langle\psi|$,
and as $\rho_i$ the reduced density matrix on site $i$, $\rho_i = \Tr_{\mathcal{L} \setminus \{i\} } |\psi\rangle\langle\psi|$, and $S(\rho) = -\Tr \rho \log \rho$.
The algorithm proceeds as follows:

\begin{algorithm}
(i) Calculate the mutual information between all pairs of sites, $I(i:j) = I(\rho_{ij}) \equiv S(\rho_i) + S(\rho_j) - S(\rho_{ij})$, and find the
pair $(i,j)$ with the largest mutual information. If all $I(i:j)$ are below some predefined threshold $\epsilon$, terminate.
(ii) Find the unitary matrix $\hat{U}_{ij}$ that acts only on sites $i$ and $j$ and maximally reduces the amount of mutual information between these
sites, i.e.\ solve $\min_{\hat{U}_{ij}} I(\hat{U}_{ij} \rho_{ij} \hat{U}_{ij}^\dagger)$.
(iii) Set $|\psi\rangle \leftarrow \hat{U}_{ij} |\psi\rangle$, and return to step 1.
\end{algorithm}
Details of the algorithm, in particular step (ii), can be found in App.~\ref{sec:disent}.
For an exact representation of a many-body state in a Hilbert space of dimension $\dim \mathcal{H}$, one iteration of the above algorithm
can be carried out with computational cost $\mathcal{O}(L \dim \mathcal{H})$ \footnote{Note that while the first iteration is carried out in $\mathcal{O}(L^2 \dim \mathcal{H})$ time, all subsequent iterations can be carried out in $\mathcal{O}(L \dim \mathcal{H})$ time, as only quantities involving the transformed sites $i$ and $j$ must be recalculated.}. For a system of non-interacting fermions, however,
the algorithm can be completely expressed in terms of the correlation matrix $C_{ij} = \langle \hat{c}_i^\dagger \hat{c}_j \rangle$~\cite{vidal2003,peschel2003,peschel2009}.
Given the
initial correlation matrix, the algorithm can be performed in $\mathcal{O}(L)$ operations per iteration, where $L$ is the number of fermionic modes. In all
cases, a single iteration of the algorithm can be performed as fast or faster than finding the eigenstates.
The number of iterations required to converge to an unentangled state depends heavily on the input state: for weakly entangled states, convergence is fast,
while for states with large entanglement, such as completely random quantum states, convergence can be very
slow. Furthermore, the algorithm is not straightforwardly applicable to certain specific classes of states (see, e.g., the perfect tensors of Ref.~\onlinecite{pastawski2015}).
We numerically explore convergence for some relevant cases in Appendix~\ref{sec:convergence}.

The algorithm ultimately constructs a unitary circuit $\hat{U} = \hat{U}_{i_\tau j_\tau}^{(\tau)} \ldots \hat{U}_{i_2 j_2}^{(2)} \hat{U}_{i_1 j_1}^{(1)}$ acting on the initial state $|\Psi\rangle$,
where $\hat{U}^{(\tau)}_{i_\tau j_\tau}$ is the unitary obtained in the $\tau$'th step.
The number of execution steps corresponds to the number of unitaries comprising the circuit. The circuit
is 2-local in the sense that each unitary acts on two sites, but it is \emph{not} local in the lattice geometry because the two sites
$i$ and $j$ may be arbitrarily far apart.
Furthermore, this circuit is not unique: an ambiguity arises since the unitary can always be followed by a swap of the two sites
or a single-site unitary while keeping the mutual information the same (see App.~\ref{sec:disent}).

\emph{Emergent geometry of unitary circuits---}
A powerful way to probe the geometry of the unitary network is to measure the length of ``geodesics'', i.e.\ the shortest paths connecting two physical
sites on the boundary of the circuit through the bulk of the circuit (see Fig.~\ref{fig:circuit}). The crucial ingredient for a numerical analysis of the unitary circuits is
an appropriate notion of length for a path in the circuit which incorporates the strength of each unitary operator. It is obvious that a careful definition
of this quantity is necessary: If, for example, one were simply to count the number of unitaries traversed in connecting two sites, one would -- for
a sufficiently deep circuit -- always find a length of 1, since eventually all pairs of sites will be directly connected by a unitary. However, deep in the circuit
the unitaries are very close to the identity, and therefore do not mediate correlations between the two sites. It is also desirable for the definition of
length to be invariant under trivial deformations of the circuit, such as introducing additional swap, identity, or single-qubit gates. Finally, the
distance measure should be computable locally and not rely on any global features of the graph.

\begin{figure}
  \centering
  \includegraphics{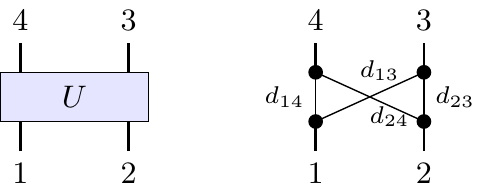}
  \caption{
  \textit{Left panel:} Labeling of the input and output indices on a unitary operator.
  \textit{Right panel:} Local graph corresponding to the unitary operator, with weights labeled on the internal edges. \label{fig:length} }
\end{figure}

Our definition of length builds on a local connection between geodesic length and
correlations~\cite{qi2013}. We construct a weighted, undirected graph as illustrated in Fig.~\ref{fig:length}: The vertices of the graph are the indices of the
unitary operators. Edges connecting different operators have weight 0, while the internal edges connecting different indices of the unitary
have lengths $d_{ab}$ as labeled in the right-hand side of Fig.~\ref{fig:length}. To define $d_{ab}$, we interpret the unitary as a wavefunction
on four qubits and set $d_{ab} = -\log[I(a:b)/ (2 \log 2)]$, where $I(a:b)$ is the mutual information between qubits $a$ and $b$ of the normalized
wavefunction. Unitarity dictates $d_{12} = d_{34} = \infty$: these two lengths are not included in the
graph. Entanglement monogamy~\cite{coffman2000,terhal2004} implies that if $d_{24}=0$ ($d_{14}=0$), $d_{13}$ ($d_{23}$) must also vanish 
and $d_{14}$ ($d_{24}$) must be infinite.
Given this weighted, undirected graph, the minimal distance between two vertices is computable using standard graph algorithms.

To develop some intuition for this quantity, consider the length of a path in well-known TNS such as MPS/PEPS and MERA~\cite{evenbly2011}.
Assuming that each tensor in such a network has roughly equal strength, we can for now simply take the length to be the number
of tensors that a path between two points traverses. For an MPS or PEPS, the length of the geodesic is then simply the
physical distance between the sites, indicative of a flat entanglement geometry.
In contrast, the length of a geodesic in a MERA scales only logarithmically with the physical distance, since the path is shorter when
moving through the bulk of the TN~\cite{evenbly2011}; this is a signature of a hyperbolic entanglement geometry.

It is important to contrast the geodesics considered here with the minimal surfaces in the RT formula for the holographic entanglement
entropy. In the standard translation to TNS, such a minimal surface is given by the minimal number of bonds that need to be cut
in order to completely separate two regions of physical sites. A minimal surface in this sense can be defined for any TN, and always
yields an upper bound to the entanglement entropy between the two regions~\footnote{This well-known fact is discussed explicitly e.g.\ in
Refs.~\onlinecite{verstraete2006,evenbly2011,cui2016quantum}.}.
While in some cases these minimal surfaces also take the form of geodesics~\cite{pastawski2015}, they are distinct from the geodesics as defined in this manuscript, which \emph{connect} pairs of sites rather than \emph{separate} regions of sites. The difference is
most easily seen in a MPS: while our geodesics are \emph{linear} in the physical distance, the minimal separating surface is \emph{constant}, since at most two bonds
need to be cut to separate the TN. While our definition is more natural
in the context of unitary circuits, they are complementary to each other, and both reveal similar information when appropriately interpreted.

It is important to recognize that while our distance measure locally is connected to correlations, there is no simple one-to-one correspondence between the behavior
of our geodesics and the behavior of two-point correlation functions. As outlined in Ref.~\onlinecite{evenbly2011}, an intuitive relation is for
correlations to decay exponentially with the geodesic length. This relation is precise for MPS, and also suggests the possibility of power-law
decay of correlations in MERA (although for certain MERA the correlations may decay faster).
However, the connection breaks down in the case of a PEPS:
while the length of a geodesic is always at least the physical (Manhattan) distance, it is possible to find PEPS whose correlations decay as a power law~\cite{verstraete2006}.
Finally, the intricate behavior in a quantum quench discussed below is largely invisible to two-point correlations.

\emph{Models---}
We first study the properties of the disentangling circuits in a model of non-interacting spinless fermions in one dimension moving in a disorder potential. We discuss further examples
in the appendix. The random-potential model is given by
\begin{equation} \label{eqn:HAnd}
\hat{H} = - t \sum_i \left( \hat{c}_i^\dagger \hat{c}_{i+1} + \hat{c}_{i+1}^\dagger \hat{c}_i \right) + \sum_i w_i \hat{c}_i^\dagger \hat{c}_i,
\end{equation}
where $\hat{c}_i^\dagger$ creates a spinless fermion on the $i$'th site of a chain of length $L$. Throughout this paper, we work with periodic boundary conditions, set $t=1$ as an overall energy scale, and focus on Slater determinants at half filling.
The random on-site potential is chosen from a uniform distribution of width $W$, $w_i \in [-W/2,W/2]$. For vanishing disorder $W \to 0$, this system is critical
and the long-wavelength limit of the ground state is described by a free-boson conformal field theory with central charge $c=1$. For any finite strength of the disorder potential,
the fermions localize~\cite{Anderson1958}. However, for very small $W \ll 1$, the localization length $\xi_\text{loc}$ is large compared to the system sizes we study, allowing us to
break translational invariance without significantly affecting physically observable properties.

\begin{figure}
  \includegraphics{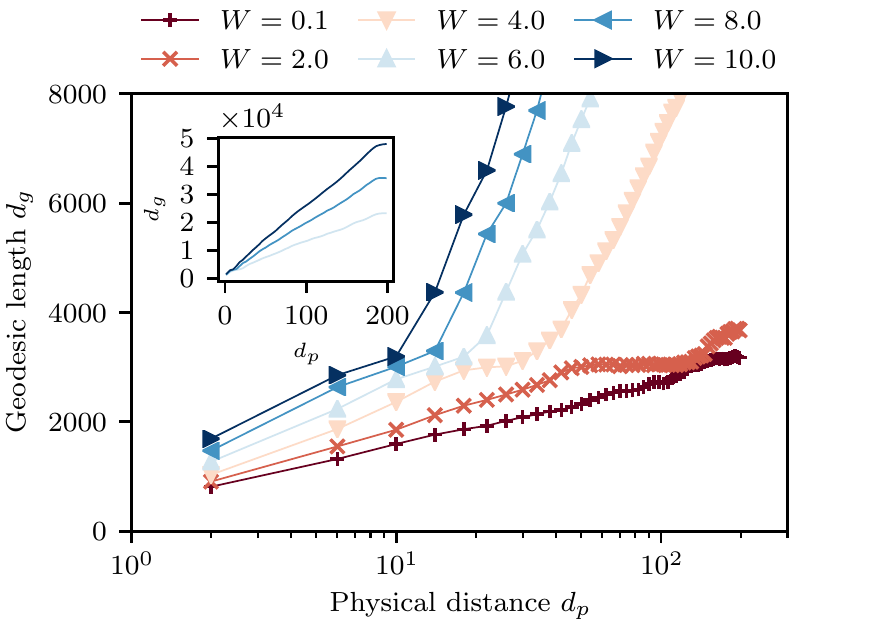}
  \caption{Geodesic length of the $L=500$ Anderson disorder model for different values of the disorder strength $W$, with 200 realizations each. While the physical distance is given in
  lattice spacings, the geodesic length is in arbitrary units. The inset shows the same data for $W=6.0,8.0,10.0$ on a linear scale to highlight the linear dependence $d_g \sim d_p$.
  \label{fig:geodesic} }
\end{figure}

\emph{Numerical results---}
Our numerical findings for the scaling with the physical distance of geodesics in ground states of~\eqnref{eqn:HAnd} are shown in Fig.~\ref{fig:geodesic} for
different disorder strengths.
Consider first the case of very large disorder strength, and thus short localization length. The geodesic length initially grows as $d_g \sim \log d_p$ with
the physical distance $d_p$ (see in particular the inset of Fig.~\ref{fig:geodesic}), and then crosses over to a linear dependence $d_g \sim d_p$, indicated by the sharp kink in Fig.~\ref{fig:geodesic}. This
behavior at large physical distance is characteristic of the flat entanglement geometry expected in a localized state.
As the disorder strength decreases, the crossover shifts to larger and larger distances, indicating that the crossover length corresponds
to the localization length. For very weak disorder potential (such as
$W=0.1$, where the localization length exceeds the system size), the region of logarithmic dependence spans the entire system.
This is the hallmark feature of hyperbolic entanglement geometry and establishes a connection to other holographic mappings, such as the AdS/CFT correspondence.

\begin{figure}
  \includegraphics[width=\columnwidth]{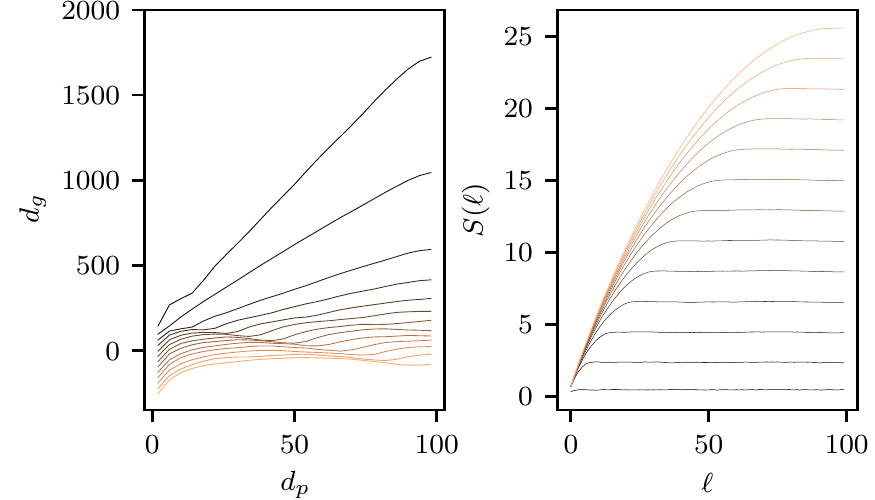}
    \caption{Quench from the ground state of the Anderson disorder model with with $L=200$ sites and $W=8$ to the clean case $W=0$. \emph{Left panel:} Geodesic length $d_g$ as a function of physical distance $d_p$. \emph{Right panel:} entanglement entropy of a contiguous region of $\ell$ sites. Individual lines represent snapshots of
  the system at times equally spaced between $T=0$ to $T=25.25$ averaged over 200 disorder realizations.
  The curves in the left panel have been offset by $-15\cdot T$. Increasing copper/decreasing blackness indicates times further in the quench. \label{fig:quench}}
\end{figure}

Going beyond eigenstates, we now consider a quench where the system is initialized in the ground state of~\eqnref{eqn:HAnd} with finite
disorder ($W=8$ in the examples chosen here), and is subsequently evolved under the translationally-invariant
Hamiltonian ($W=0$). This is similar to quenching the mass gap from a finite value to zero. We evolve up to time $T=100$, 
performing the disentangling algorithm to obtain $d_g(d_p)$ at various times during the quench.
Our results are shown in the left panel of Fig.~\ref{fig:quench}, while the right panel shows the growth
of bipartite entropy of a block of $\ell$ sites, and thus the crossover from area-law to volume-law entanglement entropy scaling. Note that here, in contrast
to Fig.~\ref{fig:geodesic}, the horizontal axis scales linearly.

Initially, the system exhibits the expected $d_g \sim d_p$ scaling of a localized system. The dominant effect at early times is a fast reduction in the scaling coefficient.
However, careful examination at early times already reveals a drastic change in the scaling behavior at short distances, where $d_g$,
instead of growing linearly with $d_p$, becomes nearly constant (or even decreases slightly). There is a sharp kink associated with the crossover from
this to the linear behavior,
which moves out to larger and larger distances with time, and finally reaches the maximal distance $d_p = L/2$. Comparison with the right panel
of Fig.~\ref{fig:quench} shows that the location of the kink corresponds to the crossover from area-law to volume-law scaling of the bipartite entanglement
entropy. Once the system has reached a long-time state with volume-law entanglement entropy, $d_g$ shows some $d_p$-dependence only for short
distances, and is flat otherwise.

In terms of the emergent entanglement geometry, the interpretation of these findings is as follows: the global quench excites
a homogeneous and finite density of local excitations, which ballistically spread and entangle with each other.
Both the kink and the area- to volume-law crossover follow the spread of this wavefront.
For distances beyond this (time-dependent) scale, the circuit is not \emph{qualitatively} affected; however, a \emph{quantitative} change in the coefficient $d_g/d_p$ occurs.
Similar to the coefficient of an area law, this quantity is easily changed by a local finite-depth unitary.
Within the characteristic length scale, on the other hand, the nature of the circuit is qualitatively changed from a short-ranged circuit
encoding an area law state to a very long-ranged circuit, with unitaries connecting the current location of an excitation to its origin,
and thus encoding volume-law entanglement. In the final state, this long-ranged circuit dominates the geodesic, with only the short-distance
behavior which originates from the boundary of the circuit exhibiting some locality. This bears resemblance to the final state in other holographic theories of
quantum quenches~\cite{hubeny2007,Abajo-Arrastia2010}, with the non-local part of the circuit playing the role of a black hole.
The relation of our results for intermediate times to the model put forward in these references is an open question left for future work.
We also note that some details of the emergent geometry, including in particular oscillations observed at times longer than the initial spreading
of entanglement shown in Fig.~\ref{fig:quench}, may be due to integrability of the model.

\emph{Outlook---}
While we have so far applied our methods to systems where a holographic description is already known, the fact that we did not make use of
any \textit{a priori} knowledge of these systems makes our methods ideally suited to systems with no known holographic
description. Most prominently, this includes the many-body localization transition~\cite{Basko06a,Oganesyan07,Pal2010,Bardarson12,bauer2013,luitz2015},
which is known to be characterized through entanglement properties~\cite{bauer2013} while the details of the transition remain
controversial~\cite{grover2014,khemani2016,yu2016}.

\begin{acknowledgments}

We thank M. P. A. Fisher, G. Refael and J. Sonner for insightful discussions, and A. Antipov and S. Fischetti for comments on earlier drafts of this manuscript.
JRG was supported by the NIST NRC Research Postdoctoral Associateship Award, by the National Science Foundation under Grant No.\ DMR-14-04230, and by the Caltech Institute of Quantum Information and Matter, an NSF Physics Frontiers Center with support of the Gordon and Betty Moore Foundation.
This material is based upon work supported by the National Science Foundation Graduate Research Fellowship under Grant No.\ DGE 1144085. Any opinion, findings, and conclusions or recommendations expressed in this material are those of the authors(s) and do not necessarily reflect the views of the National Science Foundation.
We acknowledge support from the Center for Scientific Computing at the CNSI and MRL: an NSF MRSEC (DMR-1121053) and NSF CNS-0960316.
This work was in part performed at the Aspen Center for Physics, which is supported by National Science Foundation grant PHY-1066293.
\end{acknowledgments}

%
%
%
%

\appendix

\section{Light cone growth}
\label{sec:lightcone}

\begin{figure}
  \includegraphics{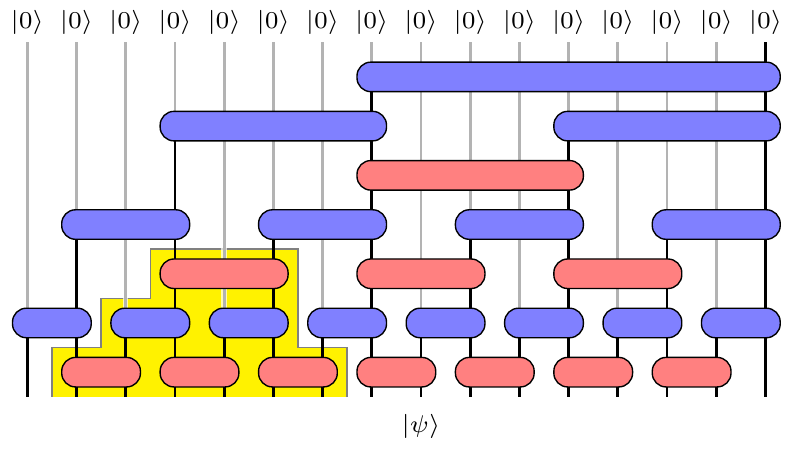}
  \caption{Two-local unitary network with the structure of a MERA state~\cite{vidal2007}, illustrating the light cone of a bulk operator.
  To make the structure of the light cone more transparent,
  the network is interpreted as acting on a product state $|0\rangle^{\otimes L}$ (at the upper end), which is evolved into the physical state $|\psi\rangle$.
  Each tensor acts only on its ends (thicker lines). The red tensors correspond to disentanglers with two input and two output qubits, while the blue
  tensors are isometries that take a qubit of an entangled state (thick line) and a previously unentangled qubit (thin line) and entangles them.
  The circuit's structure defines a light cone emanating from each unitary in the circuit.  Modification of the unitary at the top of the yellow-shaded region
  will only affect the circuit evolution and physical sites in the region; thus, the yellow region represents a light cone. \label{fig:lightcone} }
\end{figure}

In a complementary analysis to the geodesics discussed in the main manuscript, we can also characterize the emergent geometry of unitary
networks through the growth of light cones. To define the light cone, we interpret
the unitary circuit as creating the physical state from an initial product state (i.e., the reverse direction of how it is obtained in the
algorithm) and track the effect of changing one of the unitary operators. For an illustration of the light cone in a unitary circuit representing 
a MERA state, see Fig.~\ref{fig:lightcone}.
It is important to note that a notion of causality is crucial for the definition of a light cone. In the disentangling circuits, this is ensured through
the unitarity of each operator. This is a crucial difference to distance discussed in the main manuscript, which could in principle be generalized
to non-unitary networks.

We define the width of the light cone emanating from a particular
unitary operator as the number of physical sites whose state is affected by changing this unitary operator.
Quantifying the depth of the light cone, however, is more subtle. In many ansatz states, such as a scale-invariant
MERA, each layer is the same and one can thus simply count the number of layers. However,
the operators in the disentangling circuits are all different, and furthermore become closer to the
identity as the disentangling procedure progresses and the state approaches a product state. To measure the depth in the circuit,
we employ the entangling power $P(\hat{U})$; recall (from the main text) that $P(\hat{U})$ is a measure related to the amount of bipartite
entanglement a unitary can create in a multipartite state. We here use the accumulated entangling power of the steps $\tau$ up to some
step $t$,
\begin{equation}
\mathcal{P}(t) = \displaystyle\sum_{\tau < t} P(\hat{U}(\tau)),
\end{equation}
where $P(\hat{U}(\tau))$ is the entangling power of the unitary obtained in the $\tau$'th iteration of the algorithm,
to measure the depth into the circuit. We have also explored other measures for the depth, such as the total correlations~\cite{modi2010,modi2012} (see definition below), the average bipartite entropy,
and the average mutual information between pairs of sites. For all these quantities, qualitatively similar results are obtained. We restrict
our discussion to $\mathcal{P}$ since it has an interpretation purely in terms of the circuit without having to refer to the initial state
that the disentangling circuit is applied to.

In a MERA, the width of the light cone grows as $w \sim b^n$, where $b$
is the number of incoming legs on an isometry of the MERA, and $n$ is the number of layers (see Fig.~\ref{fig:lightcone}). Since the unitaries of a scale-invariant MERA
are the same in each layer, the accumulated entangling power is $\mathcal{P} \sim n$, and thus $\log (w) \sim \mathcal{P}$. This reflects the fact that the entanglement
of a critical system can be understood as a sum of equal contributions from each length scale~\cite{vidal2007}.
Indeed, since the light cone in a MERA grows as $b^n$, and the entropy of a region of size $l$ in a critical
state follows $\frac{c}{3} \log l$~\cite{srednicki1993,callan1994,holzhey1994,vidal2003}, we have that the entropy of a reduced density matrix in a MERA after $n$ layers is
$S\left( \hat{\rho}(n) \right) \sim \frac{c}{3} \log(b^n) = \frac{nc}{3} \log(b)$ and thus $S\left(\hat{\rho}(n+1)\right)-S\left (\hat{\rho}(n) \right) = \frac{c}{3} \log(b)$.
Each layer of the MERA captures the amount of entanglement encoded at a length
scale $b^n$ and makes a constant contribution proportional to the central charge of the system.

\begin{figure}
  \includegraphics{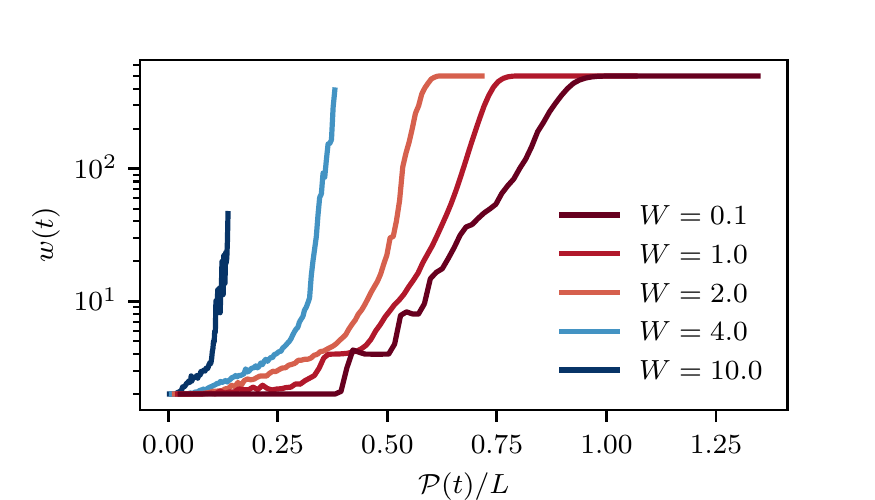}\\
  \includegraphics{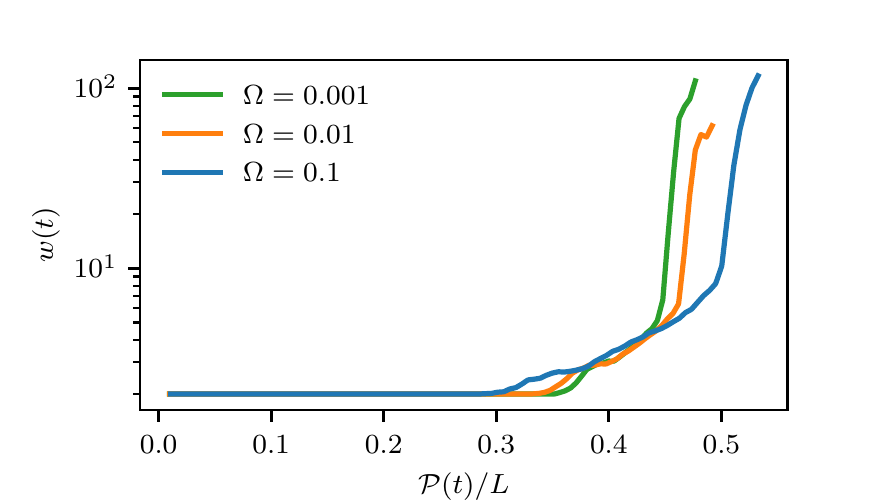}
  \caption{
  Width of the light cone $w(t)$.
  Here, $t$ represents the number of steps into the disentangling circuit.
  \textit{Top panel:} Ground states of the Anderson model for $L=500$, for 200 disorder realizations.
  \textit{Bottom panel:} Ground states in the random-singlet phase for $L=700$, for 250 disorder realization.\label{fig:lcdata1} }
\end{figure}

Below, we present numerical results for light cone growth as each step of a disentangling unitary circuit is applied.
In addition to the Anderson model, we also analyze two further examples: an analogue of a random singlet phase in a chain of free fermions, as well as
the Andr\'{e}-Aubry model of electrons in a quasi-disordered potential, which exhibits a delocalized phase even when translational symmetry
is broken.

\subsection{Anderson model}
\label{app:andersonlightcone}
Our numerical results for light cone growth in the Anderson model
\begin{equation} \label{eqn:HAnd}
\hat{H} = - t \sum_i \left( \hat{c}_i^\dagger \hat{c}_{i+1} + \hat{c}_{i+1}^\dagger \hat{c}_i \right) + \sum_i w_i \hat{c}_i^\dagger \hat{c}_i
\end{equation}
are summarized in Fig.~\ref{fig:lcdata1}. In the top panel, we show results for different strengths of
the disorder potential. For $W=0.1$, the localization length exceeds the system size and the expected scaling behavior for a critical system
is observed: after an initial regime where the light cone width
remains at $w=2$, which can be attributed to short-range non-universal physics that is encoded in local correlations, there is
a broad regime with the expected scaling of $\log(w) \sim \mathcal{P}$. This regime continues until $w$ saturates to its maximum value.
As the disorder strength in the Anderson model is increased and the localization length becomes comparable to the system size, we find that the initial
plateau becomes much shorter and the width of the light cone increases very rapidly. This is consistent with the state having
a limited amount of short-range entanglement and almost no long-range entanglement.
For very strong disorder, the final steps of the circuit exhibit a very rapid growth of $w(t)$ with $\mathcal{P}(t)$. This is due to the fact that
while almost all correlations are very local in these states,
there is a very small amount of long-range correlations which is
addressed by the last iterations of the algorithm and leads to large $w(t)$; however, since these correlations
are very weak, the unitary operators that remove them are very close to the identity and thus contribute
only very little to $\mathcal{P}(t)$. Therefore, the growth of $w(t)$ appears very steep in the final steps
of the disentangling algorithm.

\subsection{Random singlet phase}
\label{app:randomsinglet}
The Hamiltonian for the ``random singlet phase'' is given by
\begin{equation} \label{eqn:Hff}
\hat{H} = -\sum_i J_i \left( \hat{c}_i^\dagger \hat{c}_{i+1} + \hat{c}_{i+1}^\dagger \hat{c}_i \right),
\end{equation}
which we study at half filling.
For $J_i = J = 1$, this model coincides with \eqnref{eqn:HAnd} for $W=0$.
However, upon introducing disorder by choosing the $J_i$ randomly
and identically distributed, the system flows to a strongly disordered fixed point known as the random singlet phase~\cite{fisher1994}.
At the random-singlet fixed point, the low-energy states take the form of a product of maximally entangled pairs, i.e.\ for every $i$ there exists
another site $j$ such that $I(i:j) = 2 \log 2$ is maximal, while $I(i:k) = 0$ for all $k \neq j$.
Despite being very different from the ground states in the clean chain,
the entanglement scaling is similar to critical systems with an effective central charge $\tilde{c} = \log 2$~\cite{refael2004}.

To obtain the universal behavior of this fixed point in small systems, it is convenient to choose the couplings
$J_i$ from the fixed-point distribution,
\begin{equation} \label{eqn:fpdist}
P(J, \Omega) = \frac{\alpha}{\Omega} \left( \frac{\Omega}{J} \right)^{1-\alpha} \Theta(\Omega-J)
\end{equation}
where $\alpha = -1/\log \Omega$. The exponent of the distribution is controlled by $\Omega$: for $\Omega=e^{-1}\approx 0.368$, the
exponent is 0 and the distribution is a box distribution of width $\Omega$; for $\Omega \rightarrow 0$, the exponent becomes larger and larger. The random-singlet
behavior is more pronounced at short scales (high energies) for smaller $\Omega$.

The drastic difference between the structure of the random-singlet states in the lower panel of Fig.~\ref{fig:lcdata1} and the eigenstates of the Anderson
model with small $W$ in the upper panel of Fig.~\ref{fig:lcdata1} is very apparent
in the growth of the light cones.
The width of the light cone remains at $w=2$ for most of the circuit, since most of the entanglement is encoded in two-local (long-ranged)
operators. The deviations from this -- that is, the point where the light cone grows to $w > 2$ -- occur at later times as $\Omega$ is
reduced, i.e.\ the system is brought closer to the ideal random-singlet fixed point. As shown in Sec.~\ref{sec:convergence}, the disentangling
algorithm also converges drastically faster in the random-singlet phase.
Similar to the Anderson model for strong disorder, the growth of $w(t)$ is very rapid in the final stages of the algorithm.

\subsection{Andr\'{e}-Aubry model}
\label{app:andreaubry}
\begin{figure}
  \includegraphics{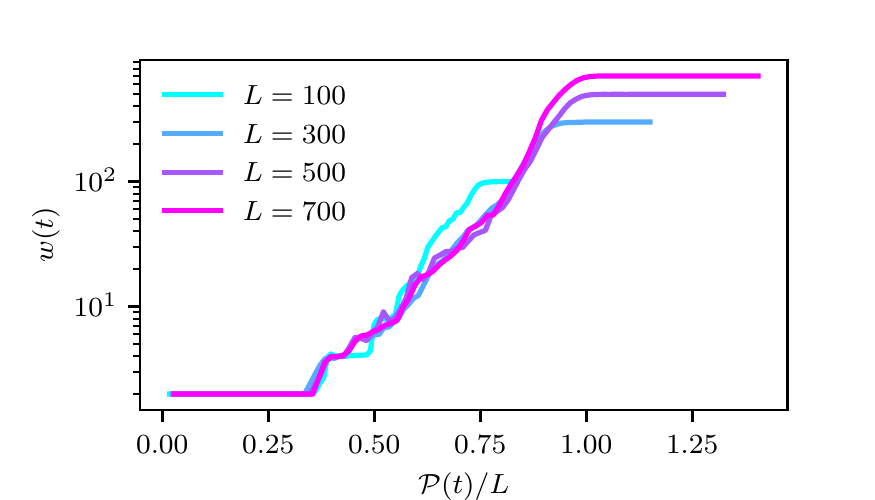}\\
  \includegraphics{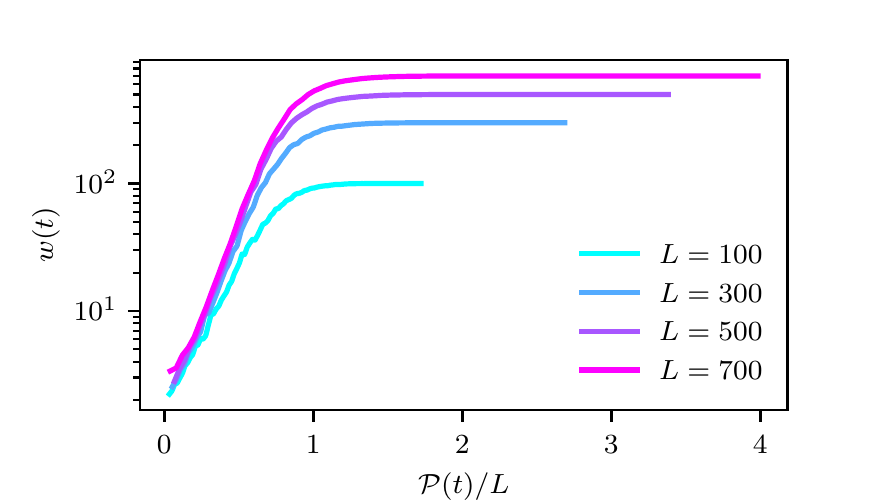}
  \caption{
      Light cone growth for the Andr\'{e}-Aubry model at $\lambda=0.1$. \emph{Top panel:} Ground states. \emph{Bottom panel:} Excited states.  Note the different scales for the axes in the two plots. \label{fig:waa} }
\end{figure}

As a final example, we consider the Andr\'{e}-Aubry model~\cite{aubry1980} given by
\begin{gather} \begin{split} \label{eqn:Haa}
\hat{H}_\text{AA} = -\sum_i\left( \hat{c}_i^\dagger \hat{c}_{i+1} + \hat{c}_{i+1}^\dagger c_i \right) \\ + \sum_i \lambda \cos(2\pi q i + \phi) \hat{n}_i ,
\end{split} \end{gather}
where $\hat{c}_i^\dagger$ creates a spinless fermion on the $i$'th site of a 1d lattice, $\hat{n}_i = \hat{c}_i^\dagger \hat{c}_i$, and we choose
$q = (\sqrt{5}+1)/2$. The model is in a delocalized regime characterized by extended wavefunctions
for $\lambda < 2$, while at $\lambda = 2$, the entire spectrum undergoes a localization transition into an Anderson insulator
for $\lambda > 2$. Upon adding interactions, the system is known to undergo a many-body localization transition~\cite{iyer2013}. The fact
that extended single-particle wavefunctions can persist even when translational invariance is locally broken by the external potential
allows us to at the same time study large systems and obtain smooth results by averaging over different choices of $\phi$.

In Fig.~\ref{fig:waa}, we show the growth of light cones in these two cases for a system at half filling. In the case of ground states for small $\lambda$, the system exhibits the expected
critical scaling $w(t) \sim \exp( \mathcal{P}(t) )$ over almost three order of magnitudes in the largest system.
In the case of excited states, on the other hand, the width of the light cones diverges very rapidly and saturates to the system size. This indicates
that the entanglement is mostly encoded globally in the state.

\section{Details of the disentangling algorithm}
\label{sec:disent}

\subsection{Ambiguity of local unitaries}

As discussed in the main manuscript, an ambiguity arises since the unitary can always be followed by a swap of the two sites
or a single-site unitary while keeping the mutual information the same.
In order to partially lift this ambiguity, we choose the unitary to minimize the
entangling power $P(\hat{U})$. To define the entangling power of a two-site unitary $\hat{U}_{ij}$, consider its decomposition
$\hat{U}_{ij} = \displaystyle\sum_\alpha \sqrt{\lambda_\alpha} \hat{X}_i^\alpha \otimes \hat{Y}_j^\alpha$, where $\hat{X}_i^\alpha$ and $\hat{Y}_i^\alpha$ are
unitary operators acting on sites $i$ and $j$, respectively, and $\Tr \left( \hat{X}_i^\alpha \hat{X}_i^\beta \right) = \Tr \left( \hat{Y}_j^\alpha \hat{Y}_j^\beta \right) = \delta_{\alpha \beta}$.
We then define $P(\hat{U}) = -\displaystyle\sum_\alpha \lambda_\alpha \log(\lambda_\alpha)$.
This quantity is closely related to the amount of entanglement the unitary can create between two systems, given additional
ancilla systems (assisted entangling power, see e.g.\ Refs.~\onlinecite{zanardi2001,wang2003}).

\subsection{Optimization of the two-site disentangling unitary}

We now describe our strategy for quickly finding a two-site unitary $U$ that maximally reduces the mutual information
between two qubits $i$ and $j$, $I(i:j) = S_i + S_j - S_{ij}$. Since the overall contribution $S_{ij}$ must remain unchanged
under unitary transformations, we only need to minimize $S_i + S_j$.

We can write a general unitary rotation on the two-qubit reduced density matrix in the form
\begin{equation}
\hat{U} = |00\rlll0| + |01\rlll1| + |10\rlll2|+|11\rlll3|,
\end{equation}
where $\ket{a} = \lbrace \ket{0}, \ldots, \ket{3} \rbrace$ is an orthonormal basis for $\mathbb{C}^4$, $\langle a|b \rangle = \delta_{ab}$. Since
we can apply a unitary to each qubit without affecting the entanglement properties, we can assume w.l.o.g.\ that $U$ is chosen such that
the reduced density matrices for the two qubits are diagonal,
\begin{eqnarray}
\rho_i &= \Tr_j \left( \hat{U} \rho \hat{U}^\dagger \right) = \left( \begin{array}{cc} p_i &0 \\ 0 &1-p_i \end{array} \right) \\
\rho_j &= \Tr_i \left(  \hat{U} \rho\hat{U}^\dagger \right) = \left( \begin{array}{cc} p_j &0 \\ 0 &1-p_j \end{array} \right),
\end{eqnarray}
where
\begin{eqnarray}
p_i &= \Tr \left[ \rho (|0\rlll0| + |1\rlll1|) \right] \\
p_j &= \Tr \left[ \rho (|0\rlll0| + |2\rlll2|) \right].
\end{eqnarray}
The quantity we want to minimize is then
\begin{equation}
S_a + S_b = H_b(p_i) + H_b(p_j),
\end{equation}
where $H_b(p) = - p \log p - (1-p) \log (1-p)$ is the binary entropy.

The strategy we pursue is to choose $\ket{a}$ to be the eigenvectors of $\rho$, in order of descending eigenvalue. If the eigenvalues of
$\rho$ are $\lambda_\alpha$, $\lambda_\alpha \geq \lambda_{\alpha+1}$, then $p_i = \lambda_0+\lambda_1$, and $p_j = \lambda_0 + \lambda_2$.
This clearly minimizes $S_i$, as well as minimizing $S_j$ under the constraint of keeping $S_i$ minimal. While we do not provide a proof
that this is the global optimum, further analytical calculation can show that this is a local minimum, and numerical tests have always shown
this to be a global minimum.

\subsection{Disentangling algorithm for free fermions}

In the case of non-interacting fermions, the entanglement properties of the system are encoded entirely in the correlation matrix (equal-time
Green's function)
\begin{equation}
    C_{kl} = \langle \hat{c}_k^\dagger \hat{c}_l \rangle,
\end{equation}
where $\hat{c}_k^\dagger$ creates a fermion on the $k$'th site of the lattice.
Given a free-fermion Hamiltonian
\begin{equation}
\hat{H} = \displaystyle\sum_{k,l} h_{kl} \hat{c}_k^\dagger \hat{c}_l = \displaystyle\sum_\alpha \epsilon_\alpha \hat{d}_\alpha^\dagger \hat{d}_\alpha,
\end{equation}
where $\epsilon_\alpha$
are the single-particle energies and $\hat{d}_\alpha^\dagger = \displaystyle\sum_i w_{\alpha i} \hat{c}_i^\dagger$ creates a fermion in the $\alpha$'th eigenstate,
\begin{equation}
C_{kl} = \displaystyle\sum_{\alpha \in F} w_{\alpha k} w^*_{\alpha l},
\end{equation}
where $F$ is the set of filled orbitals.

The entanglement of a group of sites $A$ is obtained by restricting $C_{kl}$ to the sites in $A$, $C^A_{kl} = C_{kl}$ for $k,l \in A$, computing
the eigenvalues $\lambda_\alpha$ of $C^A$, and then computing $S_A = \displaystyle\sum_\alpha H_b(\lambda_\alpha)$~\cite{vidal2003,peschel2003,peschel2009},
where $H_b$ is again the binary entropy. In particular, for a single site the entropy is simply $S_i = H_b(C_{ii})$.

The mutual information between two sites $I(i:j)$ in a free-fermion state is therefore completely encoded in the $2 \times 2$ submatrix
of the correlation matrix $C^{\lbrace ij \rbrace}$. Furthermore, if $C^{\lbrace ij \rbrace}$ is diagonal, the mutual information vanishes.
To maximally reduce the mutual information, we therefore find the orthogonal rotation $R(\theta)$ of the fermion operators that diagonalizes $C^{\lbrace ij \rbrace}$.

On the many-body operators, this transformation acts according to
\begin{align}
\hat{c}_i^\dagger &\mapsto R(\theta)_{11} \hat{c}_i^\dagger + R(\theta)_{21} \hat{c}_j^\dagger \\
\hat{c}_j^\dagger &\mapsto R(\theta)_{12} \hat{c}_i^\dagger + R(\theta)_{22} \hat{c}_j^\dagger \\
\hat{c}_i^\dagger \hat{c}_j^\dagger &\mapsto \det\left( R(\theta) \right) \hat{c}_i^\dagger \hat{c}_j^\dagger.
\end{align}
This can be succinctly summarized in the block-diagonal matrix
\begin{equation}
U = \left( \begin{array}{c|cc|c}
	1 & 0 & 0 & 0 \\ \hline
	0 & & & 0 \\
	0 &\multicolumn{2}{c|}{\smash{\raisebox{.5\normalbaselineskip}{$R(\theta)$}}} & 0 \\ \hline
	0 & 0 & 0 &\det\left( R(\theta) \right)
	\end{array} \right),
\end{equation}
where care must be taken to correctly implement fermionic anti-commutation rules when applying the off-diagonal elements.

It is interesting to note that for free fermions, the mutual information between two sites can \emph{always} be reduced to zero. While many other
entanglement properties of the free-fermion chain are similar to those of weakly interacting fermions in the same phase (for example, the CFT
description and thus the universal terms in the entanglement entropy are the same), this is a strong indication of the simpler entanglement structure
in non-interacting systems.

\subsection{Convergence}
\label{sec:convergence}

\begin{figure}
  \includegraphics{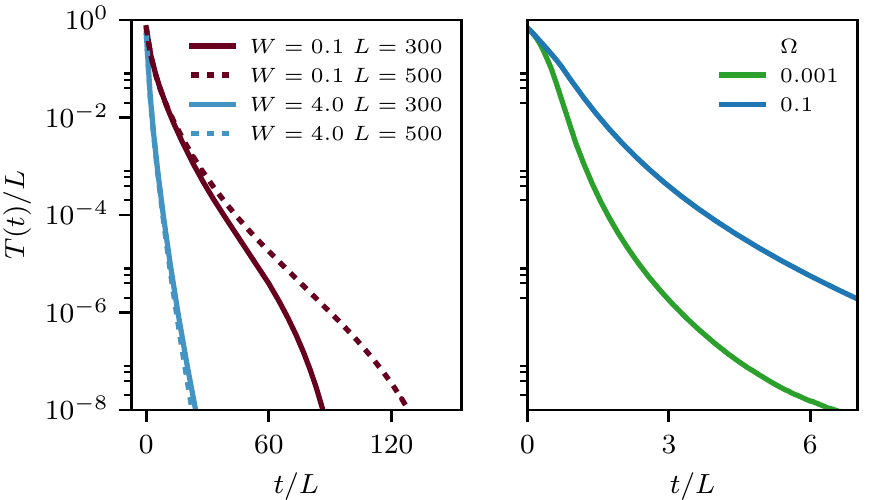}
  \caption{Convergence of the disentangling algorithm.
    Here, $T$ denotes the total correlation as given in \eqnref{eqn:totalcorr}, and $t$ represents the number of steps into the disentangling circuit.
  \textit{Left panel:} Ground states of the Anderson model~\eqnref{eqn:HAnd}. For $W=0.1$, the localization length exceeds the system size
  and the system is effectively critical on the length scales considered here, while for $W=4$ the localization length is smaller than the system size
  and the effects of localization can be observed. Averaging has been performed over 100 disorder realizations.
  \textit{Right panel:} Ground states in the random-singlet phase for $L=700$, averaged over 250 realizations. \label{fig:convergence}}
\end{figure}

To characterize the convergence of the unitary circuit towards a product state, we measure the distance from a product state~\cite{vedral1997}.
For an easily computable measure of this distance, we rely on the ``total correlation''~\cite{modi2010,modi2012}, which for a state $\hat{\rho}$ is given by
\begin{equation} \label{eqn:totalcorr}
T(\hat{\rho}) = \sum_i S(\hat{\rho}_i) - S(\hat{\rho}),
\end{equation}
where $\hat{\rho}_i$ are the reduced density matrices for sites $i$, and $S(\hat{\rho}) = - \Tr \hat{\rho} \log \hat{\rho}$.
The total correlations have the property that $T(\hat{\rho}) = \min_{\hat{\pi}} S(\hat{\rho} || \hat{\pi})$,
where $S(\hat{\rho} || \hat{\sigma})=- \Tr(\hat{\rho} \log \hat{\sigma}) - S(\hat{\rho})$ is the relative entropy, which obeys
$S(\hat{\rho} || \hat{\sigma}) \geq \vert \hat{\sigma} - \hat{\rho} \vert_1^2/2$, where $\vert \cdot \vert_1$ is the trace norm, and
$\hat{\pi} = \hat{\pi}_1 \otimes \hat{\pi}_2 \otimes \ldots \otimes \hat{\pi}_L$ is the closest product state (in relative entropy) to $\hat{\rho}$.
Since we are working over pure states,
$S(\hat{\rho}) = 0$ and $T(\hat{\rho})$ is easily computed, since we need to compute the $S(\hat{\rho}_i)$ in the course of the disentangling
algorithm. It is worth noting that for pure states $\hat{\rho}$ and $\hat{\sigma}$, $\vert \hat{\rho}-\hat{\sigma} \vert_1 = \sqrt{1-\vert \langle \hat{\rho} | \hat{\sigma} \rangle \vert^2}$.

In Fig.~\ref{fig:convergence}, we show the convergence of $T(t)/L$ (where $T(t)$ denotes the total correlation of the state after $t$ iterations
of the disentangling algorithm) for ground states of the Anderson model as well as the random-singlet
model. We consider two disorder strengths for the Anderson model: one which leads to a localization length $\xi_{loc}$ that exceeds the system size,
while the other leads to a localization length $\xi_{loc}$ short enough that localization can be observed for accessible system sizes ($L$ up to 500 sites).
In the strongly localized regime, we
find very fast convergence that is almost independent of system size. This is expected since localized states obey an area law and are known
to be generated by finite-depth local unitaries~\cite{bauer2013}. In the weakly localized regime, convergence is slower and depends much more
on the system size. This is expected since these states violate the area law with a logarithmic correction up to the relevant length scales.

For the random-singlet phase (not shown), we find very rapid convergence to a product state, even though the bipartite entanglement of the initial state is comparable to that of a critical system. This can be explained by the very simple structure of these quantum states, whose entanglement is almost all contained in simple two-site
correlations.  Convergence is faster for smaller $\Omega$, where the finite-size states are closer to the random-singlet fixed point.

\bibliography{disentangler}

\end{document}